\documentclass[preprint2]{aastex}
\begin{document}
\title{10 Years of {\em{RXTE}} Monitoring of Anomalous X-ray Pulsar
4U~0142+61: Long-Term Variability}
\author{Rim~Dib\altaffilmark{1}, Victoria~M.~Kaspi\altaffilmark{1}, and
Fotis~P.~Gavriil\altaffilmark{2}}
\altaffiltext{1}{Department of Physics, McGill University,
                 Montreal, QC H3A~2T8}
\altaffiltext{2}{X-Ray Astrophysics Laboratory, NASA Goddard Space Flight
                 Center,Greenbelt, MD 20771}
\begin{abstract}
We report on 10 yr of monitoring of the 8.7-s Anomalous X-ray Pulsar
4U~0142+61 using the {\emph{Rossi X-Ray Timing Explorer (RXTE)}}. This
pulsar exhibited stable rotation from 2000 until February 2006: the RMS
phase residual for a spin-down model which includes $\nu$, $\dot{\nu}$, and
$\ddot{\nu}$ is 2.3\%. We report a possible phase-coherent timing solution
valid over a 10-yr span extending back to March~1996. A glitch may have
occured between 1998 and 2000, but it is not required by the existing data.
We also report that the source's pulse profile has been evolving in the past
6 years, such that the dip of emission between its two peaks has been
getting shallower since 2000, almost as if the profile is recovering to its
pre-2000 morphology, in which there was no clear distinction between the
peaks. These profile variations are seen in the 2$-$4~keV band but not in
6$-$8~keV. Finally, we present the pulsed flux time series of the source in
2$-$10~keV. There is evidence of a slow but steady increase in the source's
pulsed flux since 2000. The pulsed flux variability and the narrow-band
pulse profile changes present interesting challenges to aspects of the
magnetar model.
\end{abstract}
\section{Introduction}
\label{intro}
\subsection{Anomalous X-ray Pulsars (AXPs)}
\label{intro1}
The existence of magnetars -- young, isolated neutron stars powered by the
decay of an ultrahigh magnetic field -- is now well supported by several
lines of evidence~\citep{Ref1}. There are at least two flavors of magnetars:
SGRs and AXPs. They both exhibit: X-ray pulsations with a luminosity of
10$^{34-36}$~erg~s$^{-1}$, periods ranging from 5-12~s, period derivatives
of 10$^{-13}$$-$10$^{-11}$, and surface dipolar magnetic fields of 0.6$-$7
$\times$ 10$^{14}$~G. In the magnetar model, the pulsed X-rays are the
result of a combination of surface thermal emission and resonant scattering
in the magnetosphere~\citep{Ref2}. For more on AXPs and SGRs, see reviews by
V.~Kaspi and S.~Mereghetti (this volume).
\subsection{AXP~4U~0142+61}
\label{intro2}
4U~0142+61 is an 8.7-s AXP. It has $\dot{P}$~$\cong$~0.2~$\times~10^{-11}$,
implying a surface dipole magnetic field of
1.3~$\times~10^{14}$~G\footnote{Magnetic fields here are calculated via $B
\equiv 3.2 \times 10^{19} \sqrt{P \dot{P}}$~G, where $P$ is the spin period
in seconds and $\dot{P}$ is the period derivative.}.  It is known to pulsate
in the optical band~\citep{Ref3,Ref4}, and has been detected in the
near-IR~\citep{Ref5}, in the far-IR~\citep{Ref6}, and in hard
X-rays~\citep{Ref7,Ref8}. It has a soft X-ray spectrum well fitted by a
combination of a blackbody and a power law (see, for example,
\citealt{Ref9}). AXP~4U~0142+61 rotates with high stability~\citep{Ref10}.
However, \cite{Ref11} reported a timing glitch in 1999 on the basis of an
{\em{ASCA}} observation in which the value of the frequency is marginally
discrepant with that predicted by the ephemeris reported by~\cite{Ref10}.
Here we report on continued {\em{RXTE}} monitoring observations of this
source. Our observations are described in Section~\ref{obs}. Our timing,
pulsed morphology, and pulsed flux analysis are presented, respectively, in
Sections~\ref{timing}, \ref{pulseprofile}, and~\ref{pulsedflux}. 
\section{Observations}
\label{obs}
We used 136 {\em{RXTE/PCA}} observations of various lengths in our analysis:
{\em{a)}} 4 very closely spaced {\em{RXTE}} Cycle 1 observations, {\em{b)}}
14 short Cycle 2 observations spanning a period of a year, {\em{c)}} 1 Cycle
3 observation (followed by a 2-yr gap with no observations), {\em{d)}} 118
observations taken regularly from 2000 to 2006 as part of a long-term
monitoring program spanning {\em{RXTE}} Cycles 5 to 10.  
For each observation, photon arrival times were barycentered and binned with
31.25-ms time resolution.
\section{Phase-coherent Timing}
\label{timing}
\begin{deluxetable}{lccc}
\tabletypesize{\small}
\tablewidth{385.0pt}
\tablecaption
{
Spin Parameters for 4U~0142+61\tablenotemark{a}
\label{table1}
}
\tablehead
{
&
\colhead{Pre-Gap Ephemeris\tablenotemark{b}} &
\colhead{Post-Gap Ephemeris} &
\colhead{Overall Ephemeris} \\
\colhead{Parameter} &
\colhead{Spanning} &
\colhead{Spanning} &
\colhead{Spanning} \\
&
\colhead{Cycles 1 to 3} &
\colhead{Cycles 5 to 10} &
\colhead{All Cycles}
}
\startdata
MJD range & 50170.693$-$50893.288 & 51610.636$-$53787.372 &
50170.693$-$53787.372 \\
TOAs & 19 & 118 & 137 \\
$\nu$ (Hz) & 0.115096869(3) & 0.1150969337(3) &0.1150969304(2) \\
$\dot{\nu}$ (10$^{-14}$ Hz s$^{-1}$) & $-$2.659(3) & $-$2.6935(9) &
$-$2.6514(7) \\
$\ddot{\nu}$ (10$^{-23}$ Hz s$^{-2}$) & --- & 0.417(10) & $-$1.7(2) \\
$d^{3}\nu/dt^{3}$ (10$^{-31}$ Hz s$^{-3}$) & --- & --- & 3.62(12) \\
$d^{4}\nu/dt^{4}$ (10$^{-39}$ Hz s$^{-4}$) & --- & --- & 8.7(3) \\
$d^{5}\nu/dt^{5}$ (10$^{-46}$ Hz s$^{-5}$) & --- & --- & $-$5.01(13) \\
$d^{6}\nu/dt^{6}$ (10$^{-54}$ Hz s$^{-6}$) & --- & --- & 6.6(4) \\
Epoch (MJD) & 51704.000025 & 51704.000025 & 51704.000000 \\
RMS residual & 0.019 & 0.023 & 0.019 \\
\enddata
\end{deluxetable}
\begin{figure}
\centering
\includegraphics[scale=.30]{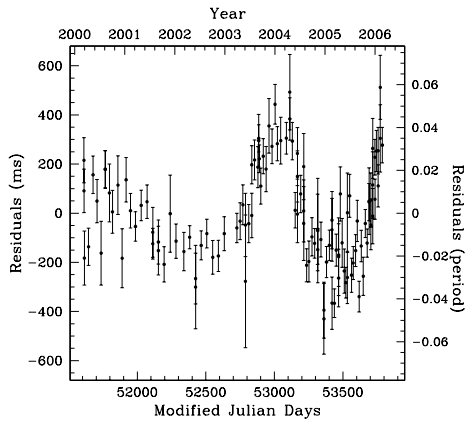}
\caption{Arrival time residuals for 4U~0142+61 for the post-gap period,
using the post-gap ephemeris given in Table~\ref{table1}.
The residuals have RMS~2.3\% of the pulse period.}
\label{figure1}
\end{figure}
\begin{figure}
\centering
\includegraphics[scale=.30]{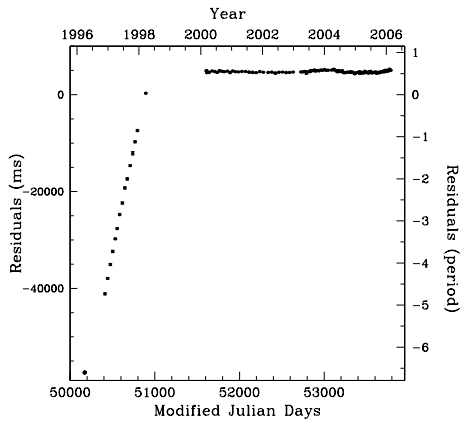}
\caption{Arrival time residuals for 4U~0142+61 for all Cycles using the
post-gap ephemeris.}
\label{figure2}
\end{figure}
\begin{figure}
\centering
\includegraphics[scale=.30]{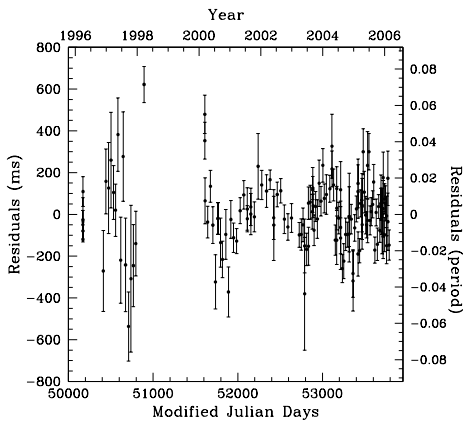}
\caption{Arrival time residuals for all Cycles using the overall ephemeris
(see Table 1).  The residuals have RMS~1.9\% of the pulse period.}
\label{figure3}
\end{figure}
Each binned time series was folded at the pulse period. Resulting pulse
profiles were cross-correlated with a high S/N template. This returned an
average pulse time of arrival (TOA) for each observation corresponding to a
fixed pulse phase. The pulse phase at any time can be expressed as a Taylor
expansion polynomial. The TOAs were fitted to the polynomial using the
pulsar timing software package
TEMPO\footnote{http://pulsar.princeton.edu/tempo.}.

We report a phase-coherent timing solution that spans the post-gap
({\emph{i.e.}} after 2000) 6-yr period up until February 2006 (MJD~53787)
including all data in {\emph{RXTE}} Cycles 5$-$10.  The parameters of our
best-fit spin-down model are shown in Table~\ref{table1}. The phase
residuals are shown in Figure~\ref{figure1}.  The best-fit post-gap
ephemeris does not, however, fit the pre-gap TOAs well. Figure~\ref{figure2}
shows a clear systematic deviation in the pre-gap residuals obtained after
subtracting the post-gap ephemeris. This could indicate that a glitch
occured at some time during the gap. However, by using 6 frequency
derivatives, we found a possible ephemeris that fits the entire Cycle~1 to
Cycle~10 range (see Table~\ref{table1}, Figure~\ref{figure3}).

The existence of our overall ephemeris cannot rule out the possibility of
the glitch having occured in 1999~\citep{Ref11}: if a fully recovered glitch
with a short relaxation time occured in the {\emph{RXTE/PCA}} observing gap
between Cycles 3 and 5, only a random phase jump would be observed. To
investigate this, we added an arbitrary but constant time jump to all the
post-gap TOAs. We were still able to find a new ephemeris that connected the
TOAs through the two-year gap. This indicates that our overall ephemeris is
not unique. Hence, we cannot rule out the possibility of a phase jump
between Cycles~3 and~5, and therefore a glitch in 1999 cannot be ruled out.
However, it is also not required by the data. Indeed, the frequency of the
discrepant {\em{ASCA}} observation reported by~\cite{Ref11} is consistent
with the frequency predicted by the overall ephemeris shown in Table~1 to
within 2$\sigma$.
\section{Pulse Profile Changes}
\label{pulseprofile}
\begin{figure}
\centering 
\includegraphics[scale=.61]{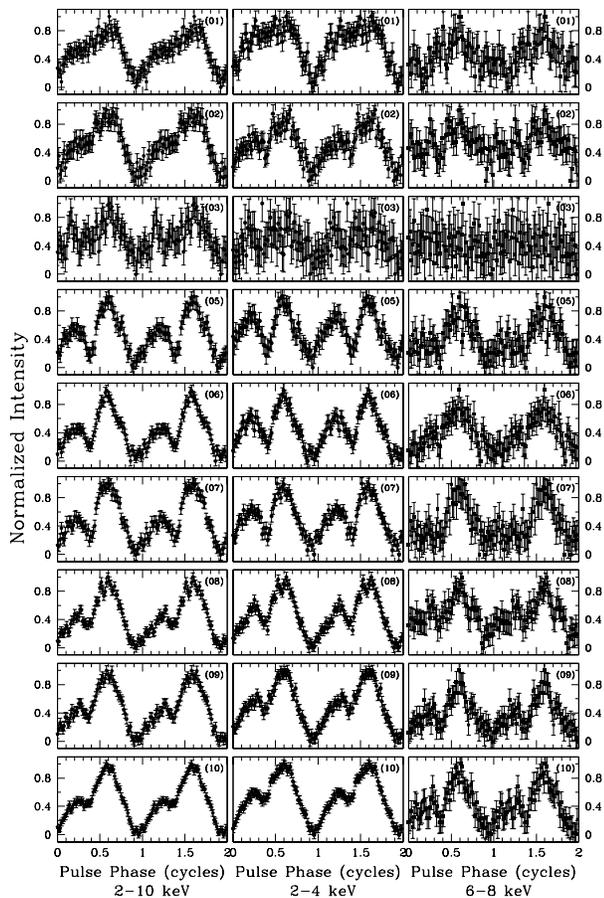}
\caption{Average pulse profiles in all RXTE Cycles in 2$-$10, 2$-$4, and
6$-$8~keV. In a given band, the different profile qualities are due to different
net exposure times. The Cycle number is shown in the top right corner of
each plot.} 
\label{figure4}
\end{figure}
\begin{figure}
\centering
\includegraphics[scale=.54]{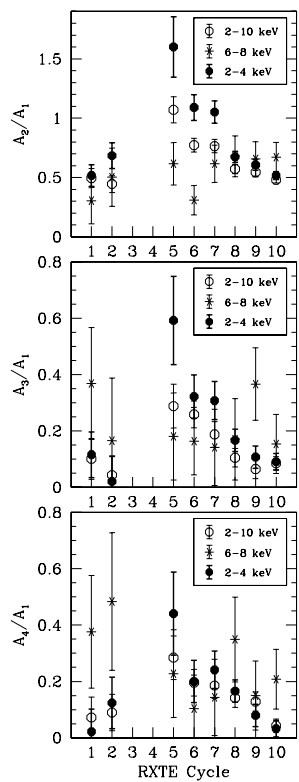}
\caption{Ratios of the Fourier amplitudes of the pulse profiles in all
three energy bands as a function of {\emph{RXTE}} Cycle. Top: ratio of the 2nd
amplitude (A2) to that of the fundamental (A1). Middle:(A3/A1). Bottom:
(A4/A1).}
\label{figure5}
\end{figure}
To search for pulse profile changes, the phase-aligned profiles were
averaged for each {\emph{RXTE}} Cycle. This was done in three energy bands.
The average profiles in all bands are presented in Figure~\ref{figure4}. The
ratios of the Fourier amplitudes of the pulse profiles in all bands as a
function of {\emph{RXTE}} Cycle are presented in Figure~\ref{figure5}.

In Figure~\ref{figure4}, the pulse profile changes are clear: for 2$-$10 and
2$-$4~keV, in Cycles~1 and~2, the smaller peak is not well defined. After
the two-year gap, in Cycle~5, the dip between the peaks is much more
pronounced. The peaks start to merge back in subsequent Cycles almost as if
the profile is recovering to its original morphology. In 6$-$8~keV, the
smaller peak is not as obvious, indicating that it has a softer spectrum
relative to the larger peak.
Given the spectrum of the source (see~\citealt{Ref9}), which is fitted to a
two-component model consisting of power law and thermal emission,
the 2$-$4~keV band includes both thermal and power-law photons while the
6$-$8~keV band contains negligibly few thermal photons. In
Figure~\ref{figure5}, the fact that the ratio of the harmonics is dropping
only in 2$-$4~keV suggests that only the thermal component of the spectrum
is evolving.
\section{Pulsed Flux}
\label{pulsedflux}
\begin{figure}
\centering
\includegraphics[scale=.30]{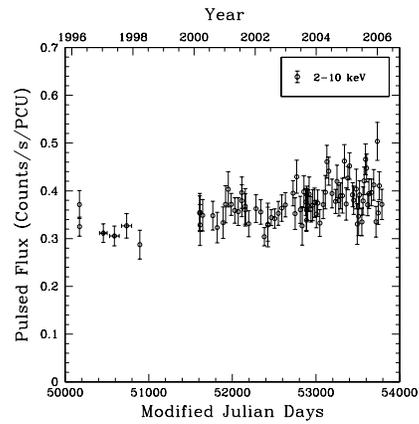}
\caption{Pulsed flux time series for 4U~0142+61 in the 2$-$10~keV band.
}
\label{figure6}
\end{figure}
For each observation, the pulsed flux, F$_{rms}$ (in counts/s/PCU), was
calculated by taking the square root of the average of the squares of the
deviations from the mean number of counts in the pulse profile. We omitted
PCU~0 from this analysis because of the uncertainties in its response due to
the loss of the propane layer.

The pulsed flux series for 4U~0142+61 shows a slow but steady increase since
2000 in 2$-$10~keV (see Fig.~\ref{figure6}). There are hints that the change
is also present in 2$-$4~keV and not in 6$-$8~keV but our statistics do not
let us confirm this. We verified there are no comparable trends in the
long-term light curves of the other AXPs observed as part of this monitoring
program.

From~\cite{Ref2}, increases in the twist angle of the field lines in the
magnetosphere can cause luminosity increases as well as pulse profile
changes. It is tempting to also attribute the increase in the pulsed flux of
this source to an increase in the twist angle. However, \cite{Ref2} also
predict that an increase in the twist angle is accompanied by an increase in
the spin down rate, which is not what we are currently observing: in the
post-gap ephemeris presented in table~1, $\ddot{\nu}$ is positive, 
i.e. the magnitude of the spin down rate is {\emph{de}}creasing.
Thus, another explanation is needed.

Finally, it is interesting to note that in April~2006, after the last
{\emph{RXTE}} Cycle included in this analysis, the pulsar appears to have
entered an extended active phase: a single burst accompanied by a pulse
profile change was detected from the pulsar on April~06~\citep{Ref12}. A series
of four bursts was later detected on June~25~\citep{Ref13}. This interesting turn
of events is presently under careful study.
\begin{acknowledgements}
This work was supported by the Natural Sciences and Engineering Research
Council (NSERC) PGSD scholarship to RD. FPG holds a National Research
Council Research Associateship Award at NASA Goddard Space Flight Center.
Additional support was provided by NSERC Discovery Grant Pgpin 228738-03
NSERC Steacie Supplement Smfsu 268264-03, FQRNT, CIAR, and CFI.  VMK is a
Canada Research Chair.
\end{acknowledgements}

\end{document}